\documentclass[]{raa}            % referee version: for submission
\usepackage{graphicx,times}
\usepackage{natbib}

\providecommand{\bm}[1]{\mbox{\boldmath$#1$\unboldmath}}

\begin{document}

   \title{Cooling of Hybrid Stars with Spin Down Compression
% $^*$
%\footnotetext{\small $*$ Supported by the National Natural Science Foundation of China.}
}

 \volnopage{ {\bf 2009} Vol.\ {\bf 9} No. {\bf XX}, 000--000}
   \setcounter{page}{1}

   \author{Kang Miao\inst{1}, Wang Xiao-Dong\inst{1}, Pan Na-Na\inst{2}
       }

   \institute{College of Physics and Electron, Henan University,
    Kaifeng, 475004, P.R.China {\it kangmiao07@gmail.com}\\
        \and
             College of Mathematics and Physics, Chongqing University of
Posts and Telecommunications, Chongqing, 400065, P.R.China\\
 \vs
\no
   {\small Received [year] [month] [day]; accepted [year] [month] [day] }
}

\abstract{We study the cooling of hybrid stars coupling with
spin-down. Due to the spin-down of hybrid stars, the interior
density continuously increases, different neutrino reactions may
be triggered(from the modified Urca process to the quark and
nucleon direct Urca process) at different stages of evolution. We
calculate the rate of neutrino emissivity of different reactions
and simulate the cooling curves of the rotational hybrid stars.
The results show the cooling curves of hybrid stars clearly depend
on magnetic field if the direct urca reactions occur during the
spin-down. Comparing the results of the rotational star model with
the transitional static model, we find the cooling behavior of
rotational model is more complicated, the temperature of star is
higher, especially when direct urca reactions appear in process of
rotation. And then we find that the predicted temperatures of some
rotating hybrid stars are compatible with the pulsar's data which
are contradiction with the results of transitional method.
\keywords{dense matter--- stars: rotation---equation of state } }

   \authorrunning{Kang Miao, Wang Xiao-Dong, Pan Na-Na }            %author_head in even pages
   \titlerunning{Cooling of Hybrid Stars with Spin Down Compression }  % title_head in odd pages
   \maketitle

%% The author head (on even pages) and the title head (on odd pages) will be
%% automatically extracted from \author{} and \title{}. Whenever the title is too long,
%% you will be asked to supply a shorter one by inserting either \authorrunning{} or
%% \titlerunning{} before \maketitle. Anyway, you can specify your own heads.
%%
%%
%% Note: In the following text body of your manuscript, please note several differences from
%%       other major journals:
%% (1) \subsection{Please Capitalize the First Letter of Each Notional Word in Subsection Title}
%% (2) Please Capitalize the First Letter of Each Notional Word in all tables' captions

%
%________________________________________________ sections below
%
\section{Introduction}           %% first-level sections will be auto-capitalized
\label{sect:intro}

The interior of neutron stars contain matter beyond the nuclear
saturation density which we have not seen yet. The cooling of
neutron stars give us an important tool to study the properties of
such dense matter. The traditional investigations of cooling often
adopt the static star model which is not connect with spin-down.
It is well known that neutron stars would spin-down due to
magnetic dipole radiation. The spin-down compression of stars may
lead to the changes in chemical composition(from nucleon matter to
deconfined quark matter) and structure(mixed phase and quark phase
appearing). The coupling of cooling and spin-down correlates
stellar surface temperatures with rotational state as well as
time. The interlinked processes of spin-down and cooling present
intriguing prospects of gain insight in the fundamental properties
of dense matter in neutron stars by confrontation with thermal
emission data from observations(Stejner et al \cite{08ste}).
 (Page et al.\cite{05pag},Yakovlev \& Pethick
\cite{04yak}).

Neutron stars are born with temperatures above $10^{10}$K. The
dominant cooling mechanism of stars is the neutrino emission from
the interior for the first several thousand years after birth,
which can be generated via numerous reactions(Yakovlev \& Pethick
\cite{04yak}). For the nucleon direct urca(NDU) reaction, the most
efficient one, is only possible if the fraction of proton exceeds
a certain threshold. It is impossible to satisfy conservation of
momentum unless the proton fraction exceed the value where both
charge neutrality and the triangle inequality
 can be observed (Lattimer et al.\cite{91lat}). Hence the traditional
 investigations of neutron stars
 often divide the cooling process into two regimes that are slow and fast cooling due to
 slow and fast neutrino emission respectively. Slow cooling occurs in the low mass stars via neutrino emission produced
 mainly by the nucleon modified Urca(NMU) process.
The fast cooling occurs in stars with mass critical one $M_{D}$(it
is of course model dependent) via the NDU process. Comparing with
the observed data, we can see that the fast cooling process would
result in contradiction between the predicted temperatures of
stars and the observations, which gives the challenge to the
traditional model.(Page et al.\cite{05pag}).

 The neutron stars containing quark matter are called hybrid stars.
Hybrid stars have more complicated interior structure and matter
composition than the purely neutron stars. Due to spin-down
compression, the interior density gradually increases, some the
thermodynamic quantities such as neutrino emission luminosity and
total heat capacity of stars continuously
 change with rotational frequency. Especially, the rate of neutrino emission
would have an
 abrupt rise because of direct urca reactions occurring which would induce the rapid fall of
temperatures of hybrid stars. Appearing of different direct Urca
processes can result in different cooling behavior in the stages
of evolution of the stars. The main difference between our model
and traditional model is that our model combine the equation of
thermal balance with the rotational stars structure and take
magnetic dipole radiation model to investigate the changes of
thermodynamic quantities and the cooling behavior of hybrid stars
with spin frequency as well as time. The simulation of cooling
curves of rotational stars are more complex than the traditional
cases because of the magnetic field dependence and changes of
rotational state. Surface temperatures of some stars including
fast cooling processes are compatible with the pulsar's data due
to spin-down.

%

%Specially, the rate of neutrino emission may have an abrupt rise
%(from NMU to NDU processes) as the interior density increases
%continuously with the spin-down of stars. Moreover, the appearance
%of quark matter would induce
% the quark direct Urca(QDU) process. Such processes are also efficient, but somewhat weaker than
% the NDU ones.
%Different neutrino processes may occur (from the modified Urca
%processes to the quark and neucleon direct Urca processes) at
%different stages of spin down which would affect the cooling
%behavior of stars.

We take Glendenning's hybrid stars model (Glendenning
\cite{97gle}) based on the perturbation
theory(Hartle\cite{67har},Chubarian et al \cite{00chu}) to study
the rotational structure of stars. In our calculation, we only
choose the simplest nucleon matter composition, namely neutrons,
protons, electrons, and muons, and ignore superfluidity and
superconductivity.

\section{Hybrid Stars}
As the stars spin-down, the nuclear matter are continuously
converted into quark matter by the exothermic reactions, i.e.
$n\rightarrow u+2d, p\rightarrow 2u+d$, s quarks immediately
appear after weak decay. The deconfinement phase transition
brought forward by Glendenning (\cite{97gle}, \cite{92gle}) who had realized
firstly the possibility of the occurrence of a mixed phase(MP) of
hadron matter and quark matter in a finite density range inside
neutron stars. Global charge neutrality of MP can be achieved by a
positively charged amount of hadronic matter and a negatively
charged amount of quark matter. The Gibbs condition for mechanical
and chemical equilibrium at zero temperature between the two
phases reads
\begin{equation}
p_{HP}(\mu_{n},\mu_{e})=p_{QP}(\mu_{n},\mu_{e})=p_{MP}.
\end{equation}
here $\mu_{n}$ is the chemical potential of neutron and $\mu_{e}$
is the  chemical potential of the electron. The condition of
global charge neutrality in the MP is
\begin{equation}
0=\frac{Q}{V}=\chi q_{QP}+(1-\chi) q_{HP}.
\end{equation}
Here $\chi=V_{Q}/V$ is the quark fraction in the MP, thus the
energy density $\epsilon_{MP}$ of the MP follows as
\begin{equation}
\epsilon_{MP}=\frac{E}{V}=\chi \epsilon_{QP}+(1-\chi)
\epsilon_{HP}.
\end{equation}
 We can obtain the equation of state (EOS) of MP using above equations. For the hadron part of star,
we adopt the Argonne $V18+\delta\upsilon+UIX^{*}$ model (Akmal et
al. \cite{98akm}) which is based on the models for the nucleon
interaction with the inclusion of a parameterized three-body force
and relativistic boost corrections. For the quark matter, we use
the EOS of a effective mass bag-model EOS(Schertle et al.
\cite{97sch}). In Figure 1, we show the EOS for hybrid stars with
deconfinement transition as described above. We choose the
parameters for quark matter EOS with s quark mass $m_{s}=150MeV$,
coupling constant $g=3$ and different bag constants
$B=108MeVfm^{-3}$ and $B=136MeVfm^{-3}$.

\begin{figure}
\centering
\includegraphics[width=80mm]{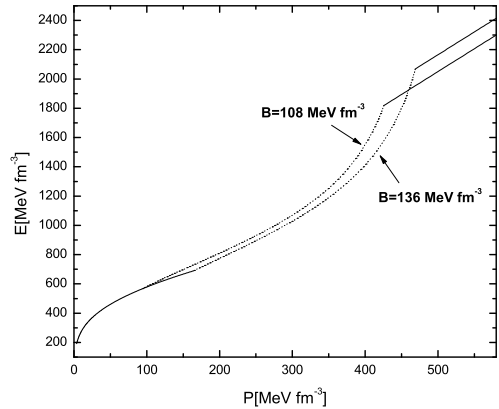}
\caption{Model EOS for the pressure of hybrid star matter as a
function of the energy density. The hadronic phase EOS is Argonne
$V18+\delta\upsilon+UIX^{*}$ model. The quark phase is effective
mass MIT bag model with bag constant B=136 and 108 $MeV fm^{-3}$
respectively. \label{fig1}}

\end{figure}
The rotation of stars lead to the change in structure. In present
work, we apply Hartle's approach (Hartle \cite{67har}) to
investigate the rotational structure of the stars. It is based on
the treatment of a rotating star as a perturbation of non-rotating
star, which can be obtained by expanding the metric of an axially
symmetric rotating star in even powers of the angular velocity
$\Omega$. We assume the frequency of stars at birth close to the
Kepler(mass-shedding limit) frequency(Hartle \& thorne
\cite{68har},Lattimer et al \cite{90lat}, De Araujo et al
\cite{95de}. We have shown the central density of rotating stars
of different gravitational mass, as a function of its rotational
frequency in Figure 2. These sequences of star models have same
total baryon number but different central density and angular
velocity. It is well known the perturbative approach fails when
the angular velocity approaches the mass-shedding limit. However,
the rotational frequencies of all known pulsars are much lower
than the mass-shedding limit. So we can use the perturbative
theory to investigate the structure of rotating stars unless we
are specifically interested in the mass-shedding problem(Benhar et
al \cite{05ben}). In Figure 2, dotted horizontal lines indicate
deconfined quark matter produced and dashed horizontal lines
indicate NDU nucleon direct Urca processes triggered. The
appearance of quark matter will result in
 the occurrence the quark direct Urca(QDU) processes. Such processes are also efficient, but somewhat weaker than
 the NDU ones.
 We observe spin-down of stars lead to the changes
in chemical composition and structure in Figure 2. In the case of
bag constant $B=108 MeV fm^{-3}$(left penal), the quark matter is
produced
 in the interior for the stars mass $M>1.4 M_{\odot}$(here denote the static mass) in process of rotation.
  For $M=1.6 M_{\odot}$ star, for example, a purely neutron star can be transformed into a hybrid star
  and  NDU reactions are triggered during spin-down. It is known that the occurrence of QDU and NDU processes
 lead to the rapid fall of temperatures of stars. In the section four, we discuss in detail spin-down
  leads to changes of the temperatures and cooling curves.
\begin{figure}
\centering
\includegraphics[width=80mm]{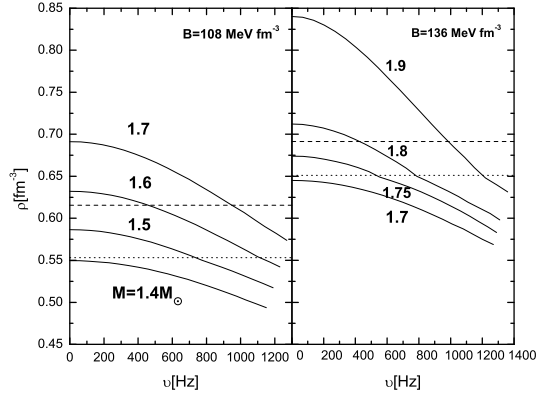}
\caption{Central density as a function of rotation frequency for
rotating hybrid stars with different gravitational mass at zero
frequency. All sequences have constant total baryon number. Dotted
horizontal lines indicate deconfined quark matter is produced and
dashed horizontal lines indicate nucleon direct Urca process is
triggered.\label{fig2}}
\end{figure}

\section{Neutrino Emissivities}
The cooling of stellar could be carried out via two channels -
neutrinos emission from the entire star and thermal emission of
photons through transport of heat from the internal layers to the
surface. The emission of neutrinos could carry away energy which
would provide efficient cooling for warm neutron stars.

For the hadron matter, we mainly consider three kinds of processes
which are NDU, NMU as well as nucleon bremsstrahlung(NB). For the
quark matter, we take QDU processes of unpaired quarks, the quark
modified Urca (QMU) processes and the quark bremsstrahlung(QB) are
considered. The most powerful neutrino emission is provided by
direct Urca process. The rate of neutrino emissivity of  NDU
processes $n\rightarrow pe\bar{\nu}$, $pe\rightarrow n\nu$ is
given by
\begin{equation}
\varepsilon_{NDU}\simeq
4.0\times10^{27}(Y_{e}\frac{\rho}{\rho_{s}})^{1/3}T_{9}^{6}\Theta_{t}~
erg cm^{-3}s^{-1}
\end{equation}
where $T_{9}$ is the temperature in units of $10^{9}$ K,
$\rho_{s}=0.16 fm^{-3}$ is the nuclear saturation density,
$\Theta_{t}=\theta(p_{Fe}+p_{Fp}-p_{Fn})$ is the threshold factor,
with $\theta(x)$ being 1 for $x>0$ and zero otherwise(Lattimer et al.\cite{91lat}). NDU
processes can occur when the fraction of proton exceeds $11\%$.
The presence of muons would raise it to about $15\%$.

The neutrino emissivity of quark matter has been given firstly by
Iwamoto\cite{82iwa}. The QDU processes have been estimated as
\begin{equation}
\varepsilon_{QDU}\simeq 8.8\times 10^{26}\alpha_c\left({\rho\over
\rho_s}\right)Y_e^{1/3}T_9^6~~\rm erg~cm^{-3}~s^{-1}
\end{equation}
with the standard value of the QCD coupling constant
$\alpha_c\simeq0.1$.

The emissivity of NMU and NB processes in the non-superfluid
$npe$ matter are usually taken from Friman\& Maxwell\cite{79fri},
in which the one-pion-exchange Born approximation with
phenomenological corrections was used for consideration. For the
emissivity of QMU and QB processes, we take the results of
Iwamoto \cite{82iwa}.

\section{Cooling Curves with Spin Down}
The traditional standard cooling model which often based on the
Tolman-Oppenheimer-Volkoff(TOV) equation of hydrostatic
equilibrium(Page et al.\cite{05pag},Yakovlev \& Pethick
\cite{04yak}). All thermodynamic quantities (neutrino emission
luminosity and total heat capacity etc)  are calculated when
rotational frequency of stars is zero.

 In present work, we investigate the cooling of hybrid stars with spin-down.
The equation of thermal balance has been assumed spherical
symmetry although it may be broken in a rotating star. It is
reasonable for slowly rotating stars which could be treated as a
perturbation to change the structure and chemical
composition(Stejner et al \cite{08ste}).
 We combine the equation of thermal balance with the rotating structure equations of the
stars(kang \& Zheng \cite{07kan}, Hartle \cite{67har}) and rewrite
the energy equation in the approximation of isothermal
interior(Glen \& Sutherland \cite{80gle})
\begin{equation}
C_{V}(T_{i},v)\frac{dT_{i}}{dt}=-L_{\nu}^{\infty}(T_{i},v)-L_{\gamma}^{\infty}(T_{s},v)
\end{equation}
\begin{equation}
C_{V}(T_{i},v)=\int_{0}^{R(v)}c(r,T)(1-\frac{2M(r)}{r})^{-1/2}4\pi
r^{2}dr
\end{equation}
\begin{equation}
 L_{\nu}^{\infty}(T_{i},v)=\int_{0}^{R(v)}\varepsilon(r,T)
(1-\frac{2M(r)}{r})^{-1/2}e^{2\Phi}4\pi r^{2}dr
\end{equation}
Where $T_{s}$ is the effective surface temperature,
$T_{i}(t)=T(r,t)$ is the red-shifted internal temperature; $T(r,t)$
is the local internal temperature of matter, and $\Phi(r)$ is the
metric function(describing gravitational red-shift)(Yakovlev \&
Haesel \cite{03yak}). Furthermore $L_{\nu}^{\infty}(T_{i},v)$and
$C_{V}(T_{i},v)$ are the total red-shifted neutrino luminosity and
the total stellar heat capacity respectively which are functions
of rotation frequency $v$ and temperature $T$; $c(r,T)$ is the heat
capacity per unit volume. $L_{\gamma}^{\infty}=4\pi R^{2}(v)\sigma
T_{s}^{4}(1-R_{g}/R)$ is the surface photon luminosity as detected
by a distant observer($R_{g}$ is the gravitational radius of
stellar ).
 The effective surface
temperature detected by a distant observer is
$T_{s}^{\infty}=T_{s}\sqrt{1-R_{g}/R}$. $T_{s}$ is obtained from
the internal temperature by assuming an envelope model(Gudmundsson
et al. \cite{83gud},Potekhin et al \cite{97pot}).The spin-down of
stars is due to the magnetic dipole radiation. The evolution of
rotation frequency is given by
\begin{equation}
\frac{dv}{dt}=-\frac{16\pi^{2}}{3Ic^{3}}\mu^{2}v^{3}\sin^{2}\theta
\end{equation}
where $I$ is the stellar moment of inertia,
$\mu=\frac{1}{2}B_{m}R^{3}$ is the magnetic dipole moment, and
$\theta$ is the inclination angle between magnetic and rotational
axes. According to the Eq.(6)-(9), we can simulate the cooling of
hybrid stars during spin-down.

%For hybrid stars, the centrifugal force decreases continuously in
%the course of which would induces the changes in stellar structure
%and chemical composition. The occurrence of different fast
%neutrino emission reactions lead to complex cooling behavior
%during spin-down of hybrid stars.

In Figure.3 we present the cooling curves of a 1.6$M_{\odot}$
rotational hybrid star for different magnetic field strengths
($10^{9}-10^{13}$G) with bag constant $B=108MeVfm^{-3}$. We find
the cooling curves of the star are different in different magnetic
fields, especially for the strong magnetic fields cases. As we
know, the stronger magnetic field is, the faster rotational
frequency slows down due to the magnetic dipole radiation, the
direct Urca processes are triggered at the earlier time, which
result in rapidly cooling in a shorter time. In the cases of
weaker fields($B<10^{10}G$), direct urca processes appear about at
the stage of photons cooling, the effect of magnetic fields is not
important at the era.
\begin{figure}
\centering
\includegraphics[width=80mm]{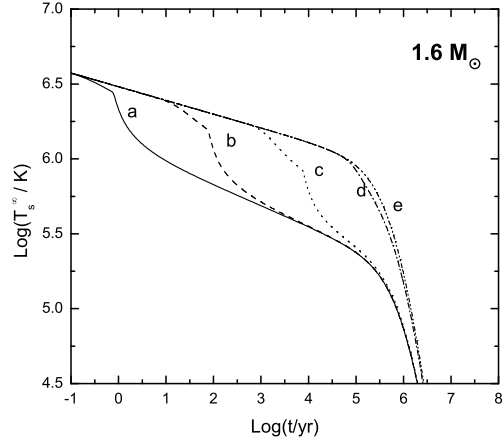}
\caption{Cooling curves of 1.6 $M_{\odot}$ rotational hybrid star
for various magnetic fields (curve a: $10^{13}$G, b: $10^{12}$G,
c: $10^{11}$G, d: $10^{10}$G, e: $10^{9}$G). \label{fig4}}
\end{figure}
 Using the same model in Figure 3, we show the neutrino
emissivity of different reactions as well as the photon luminosity
as functions of time and rotational frequency with magnetic field
$B_{m}=10^{12}G$ in Figure 4. As spin-down of the star, deconfined
quark matter appear in the core of star and then QDU reactions are
triggered at spin frequency $v=1123$Hz. From then on, the
neutrino emissivity of direct urca processes dominate the cooling
 of star. For star in age $10^{1.1}<t<10^{1.9}$yr, the QDU reactions provide most efficient neutrino emissivity.
We find NDU processes are triggered at spin frequency $v=492$Hz
which lead to the rapid increase of neutrino emissivity. The
photon emission control the cooling curves from about $t=10^{5}$yr
after birth.
\begin{figure}
\centering
\includegraphics[width=120mm]{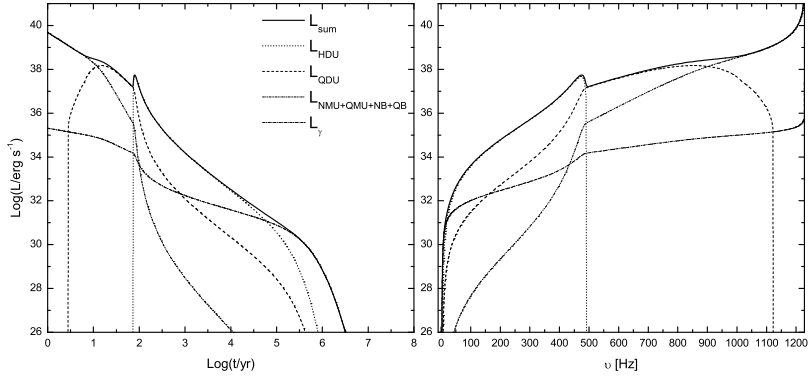}
\caption{1.6 $M_{\odot}$ hybrid star with magnetic field
$B_{m}=10^{12}$ for bag constant equal 108 $MeV fm^{-3}$, the
neutrino emissivity of different reactions as well as the photon
luminosity as functions of time(left panel) and rotation
frequency(right panel).\label{fig3}}
\end{figure}

     In Figure 5, we plot the cooling curves of hybrid stars for different mass.
      The results of the rotational model with magnetic field $B_{m}=10^{11}$G and the traditional model are both presented
     in the picture(The per star model include two cooling curves.).
      %(The curve with higher temperature correspond to the rotational model for each star).
       The results show the temperatures of the
rotating stars model are higher than the transitional static
cases, especially for the star including NDU reactions(1.6 and 1.7
$M_{\odot}$ stars for bag constant $B=108MeVfm^{-3}$, 1.8 and 1.9
$M_{\odot}$ stars for bag constant $B=136MeVfm^{-3}$).
  As we know that the traditional model use the equation of hydrostatic
equilibrium to study the configuration of stars. The central
density of stars and the thermodynamic quantities of neutrino
emission luminosity and total heat capacity etc are calculated
with rotational frequency of stars being zero. For static model,
these hybrid stars are born when direct reactions occur. For
rotational model, direct urca processes appear with the central
density gradually increases during spin-down of stars, which
induce the temperature of rotating model is higher than the static
model at the stage of neutrino cooling. From Figure 5, we can
observe the cooling curves of some rotating hybrid
stars($1.6M_{\odot}$and $1.7M_{\odot}$ in left panel,
$1.8M_{\odot}and 1.9M_{\odot}$ in right panel) are compatible with
the observed data(Page et al.\cite{04pag}, Weisskopf et
al.\cite{04wei},Slane et al.\cite{04sla}) which are contradiction
with the cooling curves of the transitional model. We find hybrid
stars have complicated cooling behavior during spin-down. For
$M=1.5 M_{\odot}$ hybrid star(left panel), the effect of spin-down
is weaker than the cases of $M=1.6 M_{\odot}$ and $1.7 M_{\odot}$
because the interior of $M=1.5 M_{\odot}$ star only appear QDU
reactions, but $M=1.6 M_{\odot}$ and $1.7 M_{\odot}$ stars include
QDU and NDU reactions in process of rotation. Comparing
$M=1.6M_{\odot}$ with $1.7 M_{\odot}$ stars(left panel), we find
earlier appearing of NDU processes lead to more rapid cooling for
the rotating hybrid stars.

\begin{figure}
\centering
\includegraphics[width=120mm]{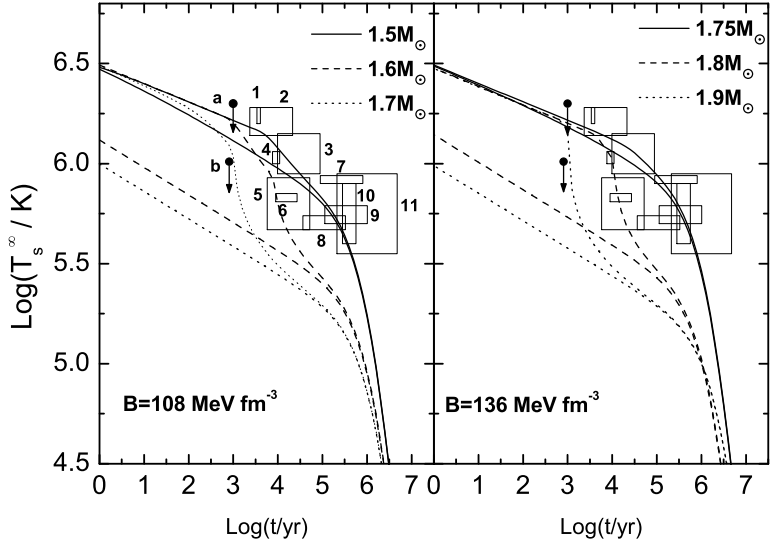}
\caption{Cooling curves of hybrid stars for different mass with
magnetic field $B_{m}=10^{11}$G and the cases in absence of
rotation.The curves of the higher temperature correspond to our
model for different mass stars.The observational data 1 to 11 are
taken from tables 1 and 2 of Page et al.\cite{04pag}. These stars
are: 1. RX J0822-4247, 2. 1E 1207.4-5209, 3. PSR 0538+2817, 4. RX
J0002+6246, 5. PSR 1706-44, 6. PSR 0833-45, 7. PSR 1055-52, 8. PSR
0656+14, 9. PSR 0633+1748, 10. RX J1856.5-3754, and 11. RX
J0720.4-3125.\label{fig5}. The next two stars, labeled as a and b:
a. PSR B0531+21(Weisskopf et al.\cite{04wei}), b. PSR
J0205+6449(Slane et al.\cite{04sla})}
\end{figure}

 We take the value of bag constant of relatively
large($B=108,136 MeV fm^{-3}$) in the paper. Comparing the results
for two parameters, we find that critical mass of hybrid stars is
smaller, the effect of spin-down is more important for the smaller
mass hybrid stars with the decrease of bag constant. In the case
of the smaller bag constant, spin-down may lead to the larger
changes for temperature of the low mass hybrid stars. However,
cooling behavior of hybrid stars for the different bag constant
are similar if the changes of the rotational state are similar
during spin-down process.

%In the case of the smaller bag constant $B=85 MeV fm^{-3}$,  it is
%important that spin-down influence the cooling behaviors for the
%hybrid stars mass $0.74M_{\odot}<M<0.88M_{\odot}$ and
%$1.05M_{\odot}<M<1.28M_{\odot}$ due to the QDU and NDU reactions
% occurring respectively during rotating evolution.
% The spin-down only lead to the
%slight change of the temperatures of hybrid stars sometimes. For
%example, the effect of rotation is not important to hybrid stars
%mass $M>1.8M_{\odot}$ for $B=108 MeV fm^{-3}$ and
%$0.88M_{\odot}<M<1.28M_{\odot},M>1.28M_{\odot}$ for $B=85 MeV
%fm^{-3}$ because the neutrino emission is not abrupt change during
%spin-down. we find the effect of spin-down is important to cooling
%of different mass hybrid stars as long as direct Urca reactions be
%triggered in process of rotation.

\section{Conclusions and Discussions}

The cooling of hybrid stars with spin-down have been studied in
the paper. We combine the cooling equation with the rotational
structure equation of stars and simulate the cooling curves of the
hybrid stars. The results show the cooling curves have a clear
magnetic field dependence if QDU or/and NDU reactions are
triggered during spin-down. The time of direct Urca reactions
triggered controls the occurring of fast cooling. Comparing the
cooling curves of the rotational models with the static models, we
find the cooling behavior of rotational models are more
complicated, the temperature of stars are higher, especially when
direct urca reactions appear in process of rotation. We also find
that the cooling curves of some rotational hybrid stars are
consistent with the observational data in case of direct Urca
neutrino emission. Through considering the inclusions of
superfluidity and superconductivity, we expect cooling curves of
these rotational hybrid stars can match the pulsar data well in
future study.
%Using Hartle's perturbative approach, we
%have investigated the structure evolution of rotating hybrid
%stars(hartle\cite{67har}).

 It has been studied by many investigators that different
EOS of hadron phase and model parameters of quark phase (bag
constant $B$, coupling constant $g$) would influence the phase
transition densities, rotational structure of hybrid stars and
corresponding internal structure etc(Schertler et al.\cite{00sch};
Pan et al.\cite{06pan}. We will investigate the effect of these
parameters on the cooling of rotating hybrid stars in detail.

% \textbf{ From the results above, we believe that the rotation
%of stars should not be ignored in studying the cooling of hybrid
%stars. For the investigation of massive stars, the spin-down may
%lead to the temperatures of stars are compatible with
%observational data which are contradiction with pulsar data for
%transitional model. }

 This work is supported by NFSC under Grant Nos.10747126.

\end{document}